\documentclass[11pt,a4paper]{article}
\usepackage{amsmath, amssymb, verbatim, graphicx, color,bbm}
\usepackage{fullpage}

\def\av{\mathbb{E}}
\def\e{\mathrm{e}}

\begin{document}

\title{Device-independent quantum key distribution\\
secure against adversaries with no long-term quantum memory}
\author{S. Pironio$^{1}$, Ll. Masanes$^{2}$, A. Leverrier$^{3,4}$, A. Ac\'{\i}n$^{2,5}$\\[0.5em]
$^{1}$ Laboratoire d'Information Quantique, Universit\'e Libre de Bruxelles (ULB),\\ 1050 Bruxelles, Belgium\\
$^{2}$ ICFO-Institut de Ciencies Fotoniques, 08860 Castelldefels, Barcelona, Spain\\
$^{3}$ Institute for Theoretical Physics, ETH Zurich, 8093 Zurich, Switzerland\\
$^{4}$ INRIA Rocquencourt, Domaine de Voluceau, B.P. 105, 78153 Le Chesnay Cedex, France\\
$^{5}$ ICREA-Instituci\'o Catalana de Recerca i Estudis Avan\c
cats, 08010 Barcelona, Spain }

\date{}

\maketitle

\begin{abstract}
Device-Independent Quantum Key Distribution (DIQKD) is a formalism
that supersedes traditional quantum key distribution, as its
security does not rely on any detailed modelling of the internal
working of the devices. This strong form of security is possible
only using devices producing correlations that violate a Bell
inequality. Full security proofs of DIQKD have been recently
reported, but they tolerate zero or small amounts of noise and are
restricted to protocols based on specific Bell inequalities. Here,
we provide a security proof of DIQKD that is both more efficient
and noise resistant, and also more general as it applies to
protocols based on arbitrary Bell inequalities and can be adapted
to cover supra-quantum eavesdroppers limited by the no-signalling
principle only. It requires, however, the extra assumption that
the adversary does not have a long-term quantum memory, a
condition that is not a limitation at present since the best
existing quantum memories have very short coherence times.
\end{abstract}

\section{Introduction}

Quantum key distribution is the art of distilling a secret key
between two distant parties, Alice and Bob, who have access to an
untrusted quantum channel \cite{RMP}. In this scenario, one
typically assumes that the equipment in Alice and Bob's labs can
be trusted, and moreover, that its behavior is accurately
described by a given theoretical model. Unfortunately, this often
turns out to be a very strong assumption which is not justified in
practice \cite{scarani}. In particular, many loopholes can be
exploited by an eavesdropper to get around the usual security
proofs: for instance, the state preparation might be imperfect
\cite{lo}, or the eavesdropper might perform a blinding attack to
take control of the detectors at a distance \cite{makarov}.

One way around such problems
consists in exhaustively listing all the potential mismatches
between the theoretical model and the real implementation and
taking care of each one of them individually. However, this approach is dubious as it is impossible to be sure that all loopholes have really been
addressed. Another, more promising, approach is inspired by the
recent framework of device-independent quantum information
processing~\cite{acinprl,Mayers}. Here, the idea is that if Alice and Bob
are able to experimentally violate a Bell inequality~\cite{Bell},
it means that their data exhibit intrinsic randomness as well as
secrecy~\cite{Ekert,BHK05}, independently of the internal
operation of the devices~\cite{acinprl}. In the recent years, this
framework has been used to prove the security of
device-independent key distribution
\cite{masanes,nonfinished,MPA,hr,BCK12,RUV12a,RUV12b,VV12}, to
certify randomness expansion \cite{Colbeck,RND,PM,Fehr,Vazirani},
self-testing of quantum computers \cite{Magniez} and
states~\cite{singlet,MYS12}, and guarantee the presence of
entanglement \cite{bancal}.

In the present work, we focus on the cryptographic task of key
distribution, which has been the subject of many very recent
developments. Until recently, security proofs were restricted to
scenarios where Alice and Bob have access to a pair of memoryless devices or $n$ independent pairs of devices, thus ensuring that the measurements inside their own labs were causally disconnected~\cite{masanes} or
commuting~\cite{MPA,hr}. This is reminiscent of the notion of
collective attacks in standard QKD, where some independence assumption is required. Ideally, one would like a protocol where
only one device is required per party, and for which no assumption is
needed for the device. This is indeed the motivation for doing
device-independent cryptography in the first place.

Recent works have been able to get rid of this assumption. In
Ref.~\cite{BCK12}, the authors introduced a protocol based on the
chained Bell inequality \cite{BC} and established its security
against arbitrary adversaries. The protocol, however, only
produces a single secret bit and does not tolerate any noise. In
Ref.~\cite{RUV12a,RUV12b}, the authors proved a strong converse of
Tsirelson's optimality result for the Clauser-Horne-Shimony-Holt
(CHSH) game, based on the CHSH inequality~\cite{CHSH}: the only
way using quantum resources to win the game as predicted by
Tsirelson's bound is to use a strategy close to the optimal one
for independent and identically distributed states, that is,
applying the optimal measurements on copies of a two-qubit
maximally entangled state. This theorem provides a security proof
for DIQKD based on the CHSH inequality. Unfortunately, the
security proof does not resistant any constant amount of noise.
While this work was completed, Vazirani and Vidick gave a
universally composable security proof of DIQKD against arbitrary
attacks \cite{VV12}. Their protocol, based again on the CHSH
inequality, is both reasonably efficient (the key length scales
linearly with the number of measurements) and tolerant to a
constant fraction of noise. A drawback, however, is that the
maximum amount of noise tolerated is of the order of $1\%$,
significantly lower than the bounds obtained for protocols using
$n$ pairs of devices.

In the present paper, we present a security proof that (i) works
for only two devices, that is, does not require commuting
measurements or memoryless devices, (ii) can be applied to generic
DIQKD protocols based on arbitrary Bell inequalities, (iii) has
the same efficiency and tolerance to noise than previous proofs
using memoryless devices. All these nice properties, however, come
at the price of assuming that the adversary only holds classical
information. While this may seem a strong requirement, it can be
easily enforced in any realistic implementation by delaying the
reconciliation process, since the best existing quantum memories
have very short coherence times~\cite{notecoh}. Another advantage
of our general framework is that it can also provide security
beyond quantum theory, that is, against eavesdroppers that are
only limited by the no-signalling principle.

The outline of the paper is the following. We first give a brief
reminder of the relation between non locality, that is, violation
of a Bell inequality, and randomness. We then describe the quantum
key distribution protocol and present its secret key rate. We
prove the security of the protocol under the assumption that the
eavesdropper does not have access to a long-term quantum memory.
We conclude by briefly comparing our results with the existing
security proofs, and discussing some rather natural follow-up
questions.

\section{Nonlocality and randomness}
In the following, we consider a bipartite scenario where Alice and
Bob input random variables $X$ and $Y$ in their respective devices and
obtain classical outputs $A$ and $B$, respectively. We denote
$\lambda_A, \lambda_B, \lambda_X, \lambda_Y$ the sizes of the
alphabets of $A,B,X,Y$, respectively. Moreover, we denote by
$P(a,b|x,y)$ the probability of getting the specific results $A=a,B=b$ when the
inputs are $X=x,Y=y$, and $P(A,B|X,Y)$ the vector with components
$P(a,b|x,y)$.

A Bell inequality can be written as
\begin{equation}\label{Bell ineq}
    I[P(A,B|X,Y)] := \sum_{a,b,x,y} \beta (a,b,x,y)\, P(a,b|x,y) \leq I_\mathrm{cl}\ ,
\end{equation}
where $I_{\rm cl}$ is the classical upper-bound. To any such Bell inequality, one can associate a bound on the randomness of the output $A$ given the input $X=x$ through a function $\tau_x$ such that
\begin{equation}\label{tau I}
    P(a|x) \leq \tau_x(I[P(A,B|X,Y)])\quad \mbox{ for all } a\in\lambda_A\ .
\end{equation}
Such a function can be computed using the techniques
given in~\cite{Q set}, as explained in~\cite{RND}. Without loss of generality, this function can be assumed to be monotonically non-increasing and such that $-\log(\tau_x(\cdot))$ is convex.

For simplicity, we consider the case where there exist an input-independent bound, i.e. a function $\tau$ such that $\tau(I)=\tau_x(I)$ for all $x\in\lambda_X$. Examples of Bell inequalities satisfying this property are: the
CHSH inequality \cite{CHSH}, the chained inequality \cite{BC}, and
the Collins-Gisin-Linden-Massar-Popescu (CGLMP) inequality
\cite{CGLMP}. Our results, however, can easily be generalised to cover the case of input-dependent bounds.

\section{Description of the protocol}
The DIQKD protocol that we consider in this paper is very general in the sense that it is compatible with arbitrary Bell inequalities, in particular with the various examples of Bell
inequalities mentioned above. Our protocol consists of four steps:
measurements, estimation of the Bell violation, error correction
and privacy amplification. We note $n$ the number of times each
device is used during the protocol.

\begin{enumerate}

\item {\bf Measurements.}
Alice and Bob respectively generate the random variables $U_j,V_j
\in \{0,1\}$ with distribution $\Pr\{U_j=1\}= \Pr\{V_j=1\}= q
=n^{-1/8}$ for $j=1,\ldots n$. If $U_j=0$ then Alice measures
round $j$ with input $0$ obtaining outcome $A_j$. If $U_j=1$ then
Alice generates $X_j$ with uniform distribution $P(x_j)
=1/\lambda_X$ and measures round $j$ with input $X_j$ obtaining
outcome $A_j$. Bob does the analog with $V_j$, input $Y_j$, and
outcome $B_j$. In other words, events where $U_j=V_j=0$ are used
to establish a raw key, while events where $U_j=V_j=1$ are used to
test the Bell inequality and guarantee that a secret key can
indeed be extracted from the raw key.

\item {\bf Estimation.}
Alice and Bob publish $(u_j, v_j)$ for all $j$, and discard the
data corresponding to the rounds with $u_j \neq v_j$. The data
corresponding to the $m$ post-selected rounds $(u_j, a_j, b_j,
x_j, y_j)$ with $v_j=u_j$ is relabeled with the index $i=1,\ldots
m$ keeping the time order. The data corresponding to the rounds of
the set ${\cal E} := \{i\, |\,  U_i=V_i=1 \}$
is also published and used to estimate the Bell-inequality violation. More specifically, Alice and Bob can use the public data to compute the following quantity:
\begin{equation}\label{bar xi}
    I_{\rm est} := \frac{\lambda_X \lambda_Y}{|{\cal E}|} \sum_{i\in {\cal E}}
    \beta (a_i, b_i, x_i, y_i)\ ,
\end{equation}
The data of the rounds not in ${\cal E}$ constitutes the raw key
of Alice $R=(A_i)_{i\notin {\cal E}}$ and Bob $S=(B_i)_{i\notin
{\cal E}}$.

\item {\bf Error correction.}
Alice and Bob publish $n_C$ bits in order to correct Bob\rq{}s
errors $S \to S\rq{}$. For sufficiently large $n_C$, all errors
are corrected $S\rq{} = R$ with high probability. Note that some
of the published bits are used to estimate how many more bits need
to be publish for a successful error correction. For large $n$,
publishing $n_C \approx nH(A|B)$ bits is enough.  For more details
about the functioning of error correction, we refer to
\cite{Rphd}.

\item {\bf Privacy amplification.} Alice generates and publishes
a two-universal~\cite{GPA} random function $F$ which maps $R$ to
an $n_K$-bit string $K=F(R)$. The number $n_K$ depends on the
published information as
\begin{equation}\label{n_K}
    n_K := \max\! \left\{0,
        \left\lfloor -m \log_2 \tau\! \left( \frac{|{\cal E}|}{m}
    \left(n^{1/8} -1\right)^2 I_{\rm est}
    - n^{-1/8} \right) -n_C
    - 2 |{\cal E}| \log_2 (\lambda_A \lambda_B)
    -\sqrt{n} \right\rfloor \right\}\ ,
\end{equation}
where $\lfloor \gamma \rfloor$ is the largest integer not bigger
than $\gamma$. Alice and Bob then compute $(F(R), F(S\rq{}))$, obtaining
two copies of the secret key.
\end{enumerate}

Note that if the adversary holds a quantum memory, but cannot keep
it for an arbitrary long time, the honest parties should implement
the protocol in two steps: (i) they receive the quantum systems
from the source and perform the measurements, (ii) some time $T$
later they perform the rest of the protocol involving the public
communication for the estimation, error correction, and privacy
amplification. We show security under the assumption that the
adversary cannot keep a quantum memory for a time $T$. According
to current and near-future technology, this assumption can be
enforced by taking $T$ of the order of a few minutes~\cite{notecoh}.

\section{Security and efficiency}
To prove security, we will not make any assumption on the behaviour of the devices of Alice and Bob, except that they do not  broadcast information about the inputs and outputs towards the adversary (a condition without which there is no hope of ever establishing any secret). Modulo this requirement, we can even assume for simplicity that the devices have been built by the adversary. The eavesdropper could in particular hold quantum systems that are entangled with the systems in the users' devices. However, our proof of security only holds under the condition that  the eavesdropper cannot store this quantum information past the measurement step of the protocol. After this step, she should thus perform a measurement $M$ on his quantum system, which would give him some classical information $E$ about the behaviour of Alice's and Bob's devices. But since until this point no public communication has been exchanged between Alice and Bob, we can as well assume that the eavesdropper has performed his measurement before the users received their devices from the source. The fact that our proof of security holds independently of the behaviour of the devices, then implies that it holds independently of the prior classical information $E$ that Eve holds on the devices, and we can thus forget $E$ in the following.

At the end of the protocol, Alice holds the secret key $K$, and Eve holds the information published in the estimation step $W=[(U_1, \ldots U_m), (A_i, B_i, X_i, Y_i)_{i\in {\cal E}}]$, in the error correction step $C= \theta(R)$, and in the privacy amplification step $F$.
Note that here we consider the worst case, where all the messages published within the error-correction step are a function $\theta$ of Alice\rq{}s raw key $R$. Let $P(k,f,w,c)$ be the probability distribution for these random variables.

We say that $K$ is an ideal secret key if it is uniformly distributed and uncorrelated with all the rest:
\begin{equation}
P(k,f,w,c)= 2^{-n_K (w)} P(f,w,c)\quad \text{for all} \quad k,f,w,c.
\end{equation}
Note that since ${\cal E}$ and $I_{\rm est}$ are functions of $w$, so is $n_K$.
It is unrealistic to expect that a protocol can generate an ideal
secret key. Instead, what we demand is that the distribution
generated by the above protocol is indistinguishable from an ideal
secret key. It is known that the optimal success probability when
discriminating the two distributions is \cite{Rphd}
\begin{equation}\label{P succ}
    p_{\rm succ} = \frac 1 2 + \frac 1 4 \sum_{k,f, w,c}
    \Big| P(k,f, w,c)- 2^{n_K (w)} P(f,w,c) \Big| \ .
\end{equation}
The main result of this work (see the Theorem below) is to shows that \begin{equation}\label{main result}
    p_{\rm succ} \leq
    \frac 1 2 + \gamma\, \e^{-\beta_0^2\, n^{1/8}}\ ,
\end{equation}
where $\gamma$ is a constant and $\beta_0 = \sqrt{8}\, \lambda_X \lambda_Y \max_{a,b,x,y} |\beta (a,b,x,y)|$. For large $n$, the success probability (\ref{main result}) tends to 1/2, which makes the optimal discriminating strategy not better than a random guess.

Let us now discuss the efficiency of the protocol in the asymptotic limit where $n$ tends to
infinity.
For large $n$ one expects,
\begin{eqnarray}\nonumber
    m &\approx& n\Pr\{ U=V \} \approx n-2n^{7/8}\ ,
\\ \nonumber
    |{\cal E}| &\approx& n\Pr\{U=V=1 \} \approx n^{3/4}\ ,
\end{eqnarray}
with high probability. This gives an asymptotic secret key rate of
\begin{equation}\label{rate}
    \lim_{n\to \infty} \frac{n_K}{n} \ =\ \log \frac{1}{\tau\! \left( I_{\rm est}  \right)}
    -H(A|B)\ .
\end{equation}
This is the same rate as the one given in \cite{MPA} for memoryless devices but with security against full quantum adversaries.

\begin{figure}
\centering
\includegraphics[scale=0.80]{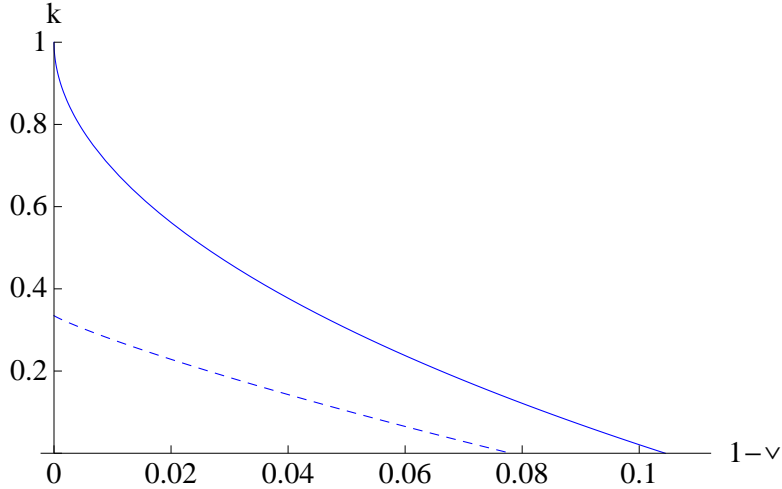}
\caption{Asymptotic secret key rate $k$ \emph{vs} noise $1-\nu$ for the CHSH protocol and a state $\rho_\nu = \nu |\phi\rangle \langle \phi| + (1-\nu) \mathbbm{1}/4$, where $|\psi\rangle$ is maximally entangled. The upper curve corresponds to a quantum adversary while the lower one considers an adversary only limited by the no-signalling principle. } \label{key}
\end{figure}

In the case of the CHSH inequality, $\beta(a,b,x,y,) = (-1)^{a \oplus b \oplus x\cdot y}$, we define $\tau_\mathrm{QM}$ and $\tau_\mathrm{NS}$ such that $p(a|x)  \leq  \tau_\mathrm{QM}(I[P(A,B|X,Y)])$ holds against an adversary limited by quantum theory and  $p(a|x)  \leq  \tau_\mathrm{NS}(I[P(A,B|X,Y)])$ holds against an adversary limited by the no-signalling principle. The specific values of these functions was derived in \cite{RND,MPA}:
\begin{align}
\tau_\mathrm{QM}(I) &=\frac{1}{2} \left(1+\sqrt{2 -\frac{I^2}{4}}\right),\\
\tau_\mathrm{NS}(I) &=\frac{1}{4} - \frac{I}{4}.
\end{align}
In Fig.~\ref{key}, we plot the asymptotic secret key rate as a function of the visibility of the state $\rho_\nu = \nu |\phi\rangle \langle \phi| + (1-\nu) \mathbbm{1}/4$ shared by Alice and Bob.

\section{Proof}
We now proceed with a detailed security proof for the protocol described above. Before we present and prove our main result which is an explicit bound on $p_\mathrm{succ}$, we need three technical lemmas.

Let us introduce a more compact notation
\begin{eqnarray}\label{def t}
    t_i &:=& \left\{
    \begin{array}{ll}
        a_i & \mbox{ if }\ i\notin {\cal E} \\
        (a_i, b_i) & \mbox{ if }\ i\in {\cal E}
    \end{array}
    \right. ,
\\ \label{def z}
    z_i &:=& \left\{
    \begin{array}{ll}
        u_i & \mbox{ if }\ i\notin {\cal E} \\
        (u_i, x_i, y_i) & \mbox{ if }\ i\in {\cal E}
    \end{array}
    \right. ,
\end{eqnarray}
for $i=1, \ldots m$. Variables with super-index $i$ represent the chain of variables associated to time steps equal or earlier than $i$, that is $t^i= (t_1, t_2, \ldots t_i)$. Recall that the information made public in the estimation step is $w= \left[ u^m, (a_i, b_i, x_i, y_i)_{i\in {\cal E}} \right]$ and that the raw key is $r=(a_i)_{i\notin {\cal E}}$. Let $g=(a_i, b_i)_{i\in {\cal E}}$ and note that $t^m = (r,g)$ and $w= (z^m, g)$.

\bigskip\noindent {\bf Lemma 1.} The no-signaling constraints imposed by the causal structure of the protocol imply
\begin{equation}
    P(t^m|z^m) \ \leq\  \tau^{m}\! \left( \bar I [t^m, z^m] \right)\ ,
\end{equation}
for all $(t^m,z^m)$, where
\begin{equation}\label{I bar}
    \bar I [t^m,z^m] := \frac{1}{m} \sum_{i=1}^m
     I\! \left[ P(A_i,B_i|X_i,Y_i, t^{i-1},z^{i-1}) \right]\ .
\end{equation}
Note that above, in $P(A_i, B_i|X_i, Y_i, t^{i-1}, z^{i-1})$, the symbols $A_i, B_i, X_i, Y_i$ are upper-case while $t^{i-1}, z^{i-1}$ are lower-case, meaning that $P(A_i, B_i|X_i, Y_i, t^{i-1}, z^{i-1})$ is the vector with components $P(a_i, b_i|x_i, y_i,t^{i-1}, z^{i-1})$ for all values of $a_i, b_i, x_i, y_i$ but fixed $t^{i-1}, z^{i-1}$.

\bigskip\noindent{\em Proof.}
This proof is based on an argument introduced in~\cite{RND}. A
useful observation is that bound~(\ref{tau I}) implies
\begin{equation}\label{extra}
    P(a,b|x,y) \leq \tau(I[P(A,B|X,Y)])\quad \mbox{ for all } a,b,x,y\ .
\end{equation}
The following chain of equalities and inequalities follows from: Bayes rule, no-signaling to the future, bounds~(\ref{tau I}) and~(\ref{extra}), and the concavity of the function $\log(\tau(\cdot))$.
\begin{eqnarray}\nonumber
    P(t^m| z^m) &=& P(t_1| z^m) P(t_2, t_3, \ldots | z^m, t_1)
\\ \nonumber &=&
    P(t_1| z^1) P(t_2, t_3, \ldots | z^m, t_1)
\\ \nonumber &=&
    \prod_{i=1}^m P(t_i| z^{i}, t^{i-1})
\\ \nonumber &\leq&
    \prod_{i=1}^m \tau\!
    \left( I\! \left[ P(A_i,B_i|X_i,Y_i, z^{i-1}, t^{i-1}) \right] \right)
\\ &\leq&
    \tau^{m}\! \left( \bar I [t^m,z^{m}] \right)
\end{eqnarray}\hfill$\square$

{\bf Lemma 2.}
The numbers $|{\cal E}|$, $I_{\rm est}$, $\bar I$ are functions of the random variable $(T^m, Z^m)$, and satisfy
\begin{equation}\label{l2}
    \Pr\! \left\{ \bar I\ \leq\
    \frac{|{\cal E}|\, I_{\rm est}}{m \Pr\{U=1|U=V\}}  - n^{-1/8}
\right\} \ \leq\
    \exp\!\left( -m\, n^{-3/4} \beta_0^{-2} \right)\ ,
\end{equation}
where $\beta_0 = \sqrt{8}\, \lambda_X \lambda_Y \max_{a,b,x,y} |\beta (a,b,x,y)|$.

\bigskip\noindent
(Here a comment is in order. Actually, $\bar I$ is not only a function of $(T^m, Z^m)$ but also depends on the global probability distribution $P(T^m, Z^m)$. But we think of this distribution as given, fixed and unknown. This dependence prevents the straight generalization of the results in this paper to a quantum adversary.)

\bigskip\noindent{\em Proof.}
The function
\[
    \eta(t,z) \ :=\ \left\{
    \begin{array}{ll}
        0 & \mbox{ if } u=0 \\
        \frac{\beta (a,b,x,y)}{P(x,y) \Pr\{U=1|U=V\}} & \mbox{ if } u=1
    \end{array}\right. ,
\]
satisfies
\begin{equation}\label{eta xi}
    \sum_{i=1}^{m} \eta[t_i, z_i] \ =\
    \frac{I_{\rm est} [t^m, z^m]\, |{\cal E}|}{\Pr\{U=1|U=V\}}\ ,
\end{equation}
and
\begin{equation}\label{new}
    \av\! \left[ \eta(T_i, Z_i) |t^{i-1}, z^{i-1} \right] =
    I[P(A_i, B_i|X_i, Y_i, t^{i-1}, z^{i-1})]\ ,
\end{equation}
for all $i$. Consider the sequence of functions of $(t^m, z^m)$ defined by
\begin{equation}\label{martingale}
    \alpha_l (t^l, z^l) \ =\ \sum_{i=1}^l \eta (t_i, z_i) - \av[\eta(T_i, Z_i)| t^{i-1}, z^{i-1}]\ ,
\end{equation}
for $l=1,\ldots m$. The fact that
\begin{equation}\label{martingale}
    \av[ \alpha_l (T^l, Z^l) | t^{l-1}, z^{l-1} ] = \alpha_{l-1} (t^{l-1}, z^{l-1})
\end{equation}
implies that the sequence of random variables $\alpha_l (T^l,
Z^l)$ is a martingale~\cite{Azuma} with respect to the sequence
$(T_l, Z_l)$. Also, using the fact that $P(x,y)= (\lambda_X
\lambda_Y)^{-1}$ and $\Pr\{U=1|U=V\} = q^2/ \left[ q^2 +
(1-q)^2\right] \geq q^2$, the differences
\begin{equation}\label{bound dif}
    |\alpha_l (t^l, z^l) - \alpha_{l-1} (t^{l-1}, z^{l-1})|
\ \leq\
    2\max_{t,z} |\eta (t,z) |
\ \leq\
    \frac{2 \max_{a,b,x,y} |\beta (a,b,x,y)|}{(\lambda_X \lambda_Y)^{-1} q^2}
\ =:\ \nu
\end{equation}
are bounded for all values of $(t^m, z^m)$. Constraints
(\ref{martingale}) and (\ref{bound dif}) constitute the premises
for Azuma\rq{}s inequality~\cite{Azuma}
\begin{equation}
    \Pr\!\left\{ \alpha_l (T^l, Z^l) \geq l\mu \right\} \leq
    \exp\!\left( \frac{-(l\mu)^2}{2\, l\, \nu^2} \right)\ ,
\end{equation}
for any $\mu >0$. Using (\ref{eta xi}), (\ref{new}) and (\ref{martingale}) we obtain
\begin{eqnarray*}
    \bar I [t^m, z^m]
&=&
    \frac{1}{m} \sum_{i=1}^m
    I\! \left[ P(A_i,B_i|X_i,Y_i,z^{i-1}, t^{i-1}) \right]
\\ &=&
    \frac{1}{m} \left(
    \sum_{i=1}^{m} \eta[t_i, z_i] - \alpha_{m} (t^{m}, z^m)
    \right)
\\ &=&
    \frac{1}{m} \left(
    \frac{|{\cal E}|\, I_{\rm est}}{\Pr\{U=1|U=V\}}
    -  \alpha_{m} (t^{m}, z^m) \right)\ ,
\end{eqnarray*}
and setting $\mu= q= n^{-1/8}$ gives (\ref{l2}). $\Box$

\bigskip\noindent {\bf Lemma 3.}
There is a good event ${\cal G}$ with probability
\begin{equation}\label{G}
    P({\cal G}) \geq
    1- 3\exp\!\left(-m\, n^{-3/4} \beta_0^{-2} \right)
    -(\lambda_A \lambda_B)^{-|{\cal E}|}\ ,
\end{equation}
such that
\begin{equation}\label{l3}
    P(r|w, {\cal G})
\ \leq\
    2\, (\lambda_A \lambda_B)^{2|{\cal E}|}\, \tau^m \!\! \left(
    \frac{|{\cal E}|\, I_{\rm est}(w)}{m \Pr\{U=1|U=V\}} -n^{-1/8} \right) ,
\end{equation}
for all $w$ such that $P(w| {\cal G}) >0$.

\bigskip\noindent {\em Proof.}
This proof uses a trick introduced in~\cite{Fehr}. The values of
$(t^m, z^m)$ in the set
\begin{equation}\label{G1}
    {\cal G}_1 := \left\{(t^m, z^m) \ \Big|\
    \bar I\ \geq\
    \frac{|{\cal E}|\, I_{\rm est}}{m \Pr\{U=1|U=V\}} -n^{-1/8}
    \right\}\ ,
\end{equation}
are the good ones, since Alice and Bob correctly lower-bound $\bar
I$ (and hence $n_K$) from the values $|{\cal E}|$ and $I_{\rm
est}$ determined in the estimation step. In the condition defining
${\cal G}_1$ above, every symbol is a constant except for $\bar I,
|{\cal E}|, I_{\rm est}$ which are functions of $(t^m, z^m)$. Note
that $\bar I$ also depends on the global distribution $P(t^m,
z^m)$, which prevents the generalization of this results to the
case of quantum adversary. Fortunately, according to Lemma 2, the
probability of ${\cal G}_1$ is large
\begin{equation}\label{PG1}
    P(\mbox{not}\, {\cal G}_1) \ <\
    \exp\!\left(-m\, n^{-3/4}\, \beta_0^{-2} \right)\ .
\end{equation}
Note the abuse of notation $P({\cal G}_1)= \Pr\{(T^m, Z^m) \in {\cal G}_1\}$. Define the set
\begin{equation}\label{G2}
    {\cal G}_2 := \left\{w \ |\
    P( {\cal G}_1 | w) \geq 1/2
    \right\}\ ,
\end{equation}
and note that $P(\mbox{not}\, {\cal G}_1 | \mbox{not}\,{\cal G}_2) > 1/2$. Using this and $P(\mbox{not}\, {\cal G}_1) \geq P(\mbox{not}\,{\cal G}_1 |\mbox{not}\,{\cal G}_2)\, P(\mbox{not}\,{\cal G}_2)$ we obtain $P(\mbox{not}\,{\cal G}_2) < 2 P(\mbox{not}\,{\cal G}_1)$.

Recall $G=(A_i, B_i)_{i\in {\cal E}}$ and note that $T^m =(R,G)$ and $W= (Z^m, G)$.
Define the set
\begin{equation}\label{G3}
    {\cal G}_3 := \left\{(g, z^m) \ \Big|\
    P(g| z^m) \geq (\lambda_A \lambda_B)^{-2|{\cal E}|}
    \right\}\ ,
\end{equation}
and note that
\begin{eqnarray}
    P(\mbox{not}\, {\cal G}_3)
\nonumber &=&
    \sum_{(g, z^m) \notin {\cal G}_3} P(z^m)\, P(g|z^m)
\\ \nonumber &<&
    \sum_{g, z^m} P(z^m)\,
    (\lambda_A \lambda_B)^{-2|{\cal E}|}
\\ &=&
    (\lambda_A \lambda_B)^{-|{\cal E}|}\ ,
\end{eqnarray}
where we have used $\sum_g 1 = (\lambda_A \lambda_B)^{|{\cal E}|}$. The good event mentioned in the statement of this lemma is ${\cal G} =$\lq\lq{}${\cal G}_1 \, \mbox{and}\, {\cal G}_2 \, \mbox{and}\, {\cal G}_3$\rq\rq{}, and has probability $P({\cal G})\geq 1- P(\mbox{not}\, {\cal G}_1) -P(\mbox{not}\, {\cal G}_2) -P(\mbox{not}\, {\cal G}_3)$, as in (\ref{G}).

We assume $(g,z^m) \in {\cal G}_2 \cap {\cal G}_3$, since it is a premise of the lemma. If $(r, g, z^m) \notin {\cal G}_1$ then $P(r|g, z^m, {\cal G}_1) =0$. Hence, the non-trivial case happens for $(r, g, z^m) \in {\cal G}_1$, which we assume in what follows. Using Bayes rule, the definition of ${\cal G}_2$ and ${\cal G}_3$, Lemma 1, and (\ref{G1}), we obtain
\begin{eqnarray}\nonumber
    P(r|g, z^m, {\cal G}_1)
&\leq&
    \frac{P(r | g, z^m)}{P({\cal G}_1|g, z^m)}
\\ \nonumber &\leq&
    \frac{P(r,g |z^m)}{P({\cal G}_1|g, z^m)\, P(g| z^m)}
\\ \nonumber &\leq&
    2\, (\lambda_A \lambda_B)^{2|{\cal E}|}\, \tau^m \!\!
    \left( \bar I [r, g, z^m] \right)
\\ &\leq&
    2\, (\lambda_A \lambda_B)^{2|{\cal E}|}\, \tau^m \!\! \left(
    \frac{|{\cal E}|\, I_{\rm est}(g, z^m)}{m\Pr\{U=1|U=V\}}
    -n^{-1/8} \right)\ ,
\end{eqnarray}
which shows the lemma. $\square$

\bigskip\noindent {\bf Theorem.} The distance between the secret key generated by the protocol and an ideal key is
\[
    \sum_{k, f, w, c} \left|
    P(k, f, w, c) - 2^{-n_K (w)} P(f, w, c)
    \right|
\ \leq\
    2^{(1- n^{1/2})/2}
    + 6\, \e^{-m\, n^{-3/4}\, \beta_0^{-2} }
    +2 (\lambda_A \lambda_B)^{-|{\cal E}|}
    \ .
\]

\bigskip\noindent {\em Proof.}
Using definitions (\ref{n_K}) and (\ref{Pguess}), Lemma 3, and $\sum_c 1 = 2^{n_C}$, we obtain:
\begin{eqnarray*}
    P_{\rm guess} (R|C; w, {\cal G})
&=&
    \sum_c \max_r P(r, c |w, {\cal G})
\\ &=&
    \sum_c \max_{\substack{r \\ \theta(r)=c}} P(r |w, {\cal G})
\\ &\leq&
    \sum_c 2\, (\lambda_A \lambda_B)^{2|{\cal E}|}\, \tau^m \!\! \left(
    \frac{|{\cal E}|\, I_{\rm est}(g)}{m \Pr\{U=1|U=V\}} -n^{-1/8} \right) ,
\\ &=&
    2^{1 - n_K(g)-\sqrt{n}} \ .
\end{eqnarray*}
The symbol $P_{\rm guess} (R|C; w, {\cal G})$ denotes the knowledge of $R$ with respect to $C$ (see Appendix) when the statistics is conditioned on the events $W=w$ and ${\cal G}$. Next, we use the identity
\begin{equation}\label{hG}
    P(t^m, z^m)= P({\cal G})  P(t^m, z^m| {\cal G})
    +P(\mbox{not}\, {\cal G})  P(t^m, z^m| \mbox{not}\, {\cal G})
\end{equation}
with the event ${\cal G}$ introduced in Lemma 3.
Noticing that $(K, F, W, C)$ is a function of $(T^m, Z^m, F)$, using (\ref{hG}), the triangular inequality, and Lemma 4, we see that
\begin{eqnarray*}
&&
    \sum_{k, f, w, c} \left| P(k, f, w, c) - 2^{-n_K (w)} P(f, w, c) \right|
\\ &\leq &
    \sum_{k, f, w, c} \left| P(k, f, w, c |{\cal G}) - 2^{-n_K (w)} P(f, w, c| {\cal G}) \right|
    + 2\, P(\mbox{not}\, {\cal G})
\\ &\leq &
    \sum_{k, f, w, c} P(w| {\cal G}) \left| P(k, f, c |w, {\cal G}) - 2^{-n_K (w)} P(f, c|w, {\cal G}) \right|
    + 2\, P(\mbox{not}\, {\cal G})
\\ &\leq &
    \sum_{w} P(w| {\cal G})\,
    \sqrt{2^{n_K (w)} P_{\rm guess} (R|H; w, {\cal G})}
    + 2\, P(\mbox{not}\, {\cal G})
\\ &\leq &
    \sum_{w} P(w| {\cal G})\,
    2^{(1- n^{1/2})/2}
    + 2\, P(\mbox{not}\, {\cal G})
\\ &= &
    2^{(1- n^{1/2})/2}
    + 6\exp\!\left(-m\, n^{-3/4}\,\beta_0^{-2} \right)
    +2 (\lambda_A \lambda_B)^{-|{\cal E}|}\ ,
\end{eqnarray*}
which concludes the proof. $\square$

\section{Conclusions}

In this work, we provide a novel security proof for DIQKD.
Contrary to most of the existing proofs, it applies to the
situation in which Alice and Bob generate the raw key using two
devices. In particular, it does not need to assume that the
devices are memoryless or, equivalently, that each raw-key symbol
is generated using a different device. While there exist other
recent proofs that also work without this assumption, they
tolerate zero~\cite{BCK12,RUV12a,RUV12b} or rather small amounts
of noise~\cite{VV12}. Another important feature of our proof is
that it can also be applied to non-signalling supra-quantum
eavesdroppers. All these advantages come at the price of making an
extra assumption on Eve: she does not have access to a long-term
quantum memory and, therefore, effectively she cannot store
quantum information. While this may at first be considered a strong
assumption (and is
actually not needed in new security proofs for
DIQKD~\cite{BCK12,RUV12a,RUV12b,VV12}), it is a very realistic
assumption taking into account current technology.

The natural open question is to understand how the assumption on
the memory can be removed within the framework presented here, or
how the other proofs~\cite{BCK12,RUV12a,RUV12b,VV12} could be
improved to tolerate realistic noise rates. In the case of
no-signalling eavesdroppers, there is some evidence suggesting
that the fact that Eve can store information and delay her
measurement prevents any form of privacy amplification between the
honest parties~\cite{nprivampl}. However, the recent results
of~\cite{VV12} imply that privacy amplification is indeed possible
against quantum eavesdroppers. A good understanding of privacy
amplification in the device-independent quantum scenario is
probably the missing ingredient to get robust and practical fully
device-independent security proofs.

\section*{Acknowledgements}
We acknowledge useful discussion with Serge Massar. This work is
supported by the ERC SG PERCENT, by the EU projects Q-Essence and
QCS, by the CHIST-ERA DIQIP project, by the Spanish  FIS2010-14830
projects, by the SNF through the National Centre of Competence in
Research ``Quantum Science and Technology'', by CatalunyaCaixa, by
the Interuniversity Attraction Poles Photonics@be Programme
(Belgian Science Policy), by the Brussels-Capital Region through a
BB2B Grant, and from the FRS-FNRS under project DIQIP.

\section*{Appendix}

A random function $F: {\cal R} \to \{0,1\}^n$ is
two-universal~\cite{GPA} if
\[
    \Pr \{ F(r)= F(r\rq{})\} \leq 2^{-n}\ ,
\]
for all $r, r\rq{} \in {\cal R}$ with $r\neq r\rq{}$. The
following is a simple extension of the main result in~\cite{GPA}.

\bigskip \noindent
{\bf Lemma 4.} Let $R,E$ be two (possibly correlated) random variables where $R$ takes values in the set ${\cal R}$, and let $F: {\cal R} \to \{0,1\}^n$ be a two-universal random function~\cite{GPA}.
The random variable $K=F(R)$ satisfies
\begin{equation}\label{lA}
    \sum_{k,f,e} \left| P(k,f,e) - 2^{-n} P(f,e) \right|
\leq
    \sqrt{2^n\, P_{\rm guess} (R|E)}\ ,
\end{equation}
where
\begin{equation}\label{Pguess}
    P_{\rm guess} (R|E)= \sum_e \max_r P(r,e)\ .
\end{equation}

\bigskip \noindent
{\em Proof. }
Using the convexity of the square function, the fact that $F$ is independent of $R,E$ and two-universality we obtain:
\begin{eqnarray*}
&&
    \left( \sum_{k,f,e} \left| P(k,f,e) - 2^{-n} P(f,e) \right| \right)^2
\\ &\leq&
    \sum_{k,f,e} P(f,e)\, 2^{-n}    \left( 2^n \sum_r P(r|e)\, \delta_{f(r)}^k -1 \right)^2
\\ &=&
    \sum_{f,e} P(f,e)\, 2^{-n}  \left(  2^{2n} \sum_{r, r\rq{}} P(r|e) P(r\rq{}|e)\, \delta_{f(r)}^{f(r\rq{})} +2^{n} -2^{1+n}\right)
\\ &=&
    -1+ 2^{n} \sum_{f,e} P(f,e) \left( \sum_{r\neq r\rq{}} P(r|e) P(r\rq{}|e)\, \delta_{f(r)}^{f(r\rq{})} + \sum_{r} P(r|e)^2 \right)
\\ &\leq&
    2^n\, \sum_e P(e) \sum_{r} P(r|e)^2
\\ &\leq&
    2^n\, P_{\rm guess} (R|E)\ .
\end{eqnarray*}\hfill$\square$



\begin{thebibliography}{99}

\bibitem{RMP}
V. Scarani, H. Bechmann-Pasquinucci, N.~J. Cerf, M. Du{\v{s}}ek,
N. L{\"u}tkenhaus and M. Peev, {\em The security of practical
quantum key distribution}, Rev. Mod. Phys. {\bf 81}, 1301 (2009).

\bibitem{scarani}
V. Scarani, and C. Kurtsiefer, {\em The black paper of quantum
cryptography: real implementation problems}, arXiv:0906.4547.

\bibitem{lo}
F. Xu,, B. Qi, H.-K. Lo, {\em Experimental demonstration of
phase-remapping attack in a practical quantum key distribution
system}, New J. Phys. {\bf 12} 113026 (2010).

\bibitem{makarov}
L. Lydersen, C. Wiechers, C. Wittmann, D. Elser, J. Skaar and V.
Makarov, {\em Hacking commercial quantum cryptography systems by
tailored bright illumination}, Nature Photonics \textbf{4} 686
(2010).

\bibitem{acinprl}
A. Acin, N. Brunner, N. Gisin, S. Massar, S. Pironio and V.
Scarani, {\em Device-independent security of quantum cryptography
against collective attacks}, Phys. Rev. Lett. \textbf{98}, 230501
(2007).

\bibitem{Mayers}
D. Mayers and A. Yao, {\em Self testing quantum apparatus}, Quant.
Inform. Comput. {\bf 4} 273 (2004).

\bibitem{Bell}
J.~S. Bell, \emph{Speakable and unspeakable in quantum mechanics},
Cambridge University Press (Cambridge, 1987).

\bibitem{Ekert}
A. Ekert, {\em Quantum cryptography based on Bell's theorem},
Phys. Rev. Lett. {\bf 67} 661 (1991).

\bibitem{BHK05}
J. Barrett, L. Hardy, and A. Kent, {\em No signaling and quantum
key distribution}, Phys. Rev. Lett. {\bf 95} 010503 (2005).

\bibitem{pironio09}
S. Pironio, A. Acin, N. Brunner, N. Gisin, S. Massar and V.
Scarani, {\em Device-independent quantum key distribution secure
against collective attacks}, New J. Phys. \textbf{11}, 045021
(2009).

\bibitem{masanes}
L. Masanes, {\em Universally composable privacy amplification from
causality constraints}, Phys. Rev. Lett. {\bf 102} 140501 (2009).

\bibitem{nonfinished}
L. Masanes, R. Renner, M. Christandl, A. Winter and J. Barrett,
{\em Unconditional security of key distribution from causality
constraints}, arXiv:quant-ph/0606049.

\bibitem{MPA} L. Masanes, S. Pironio and A. Ac{\'i}n,
{\em Secure device-independent quantum key distribution with
causally independent measurement devices}, Nature Comm. {\bf 2}
238 (2011).

\bibitem{hr}
E. H\"anggi and R. Renner, {\em Device-independent quantum key
distribution with commuting measurements}, arXiv:1009.1833.

\bibitem{BCK12}
J. Barrett, R. Colbeck and A. Kent,  {\em Unconditionally secure
device-independent quantum key distribution with only two
devices}, arXiv:1209.0435.

\bibitem{RUV12a}
B. Reichardt, F. Unger and U. Vazirani, {\em A classical leash for
a quantum system: Command of quantum systems via rigidity of CHSH
games}, arXiv:1209.0448.

\bibitem{RUV12b}
B. Reichardt, F. Unger and U. Vazirani,  {\em Classical command of
quantum systems via rigidity of CHSH games}, arXiv:1209.0449.

\bibitem{VV12}
U. Vazirani and T. Vidick, {\em Fully device independent quantum
key distribution}, arXiv:1210.1810.

\bibitem{notecoh}
The two main approaches for quantum memories are based on ensemble
of atoms or on crystals. To our knowledge, the best existing
quantum memories with ensemble of atoms have coherence times of
the order of 100 milliseconds, A. G. Radnaev \emph{et al.}, Nature
Phys. \textbf{6}, 894 (2010). Moving to crystals, coherence times
of the order of a few seconds have been reported for classical
light, see J.~J. Longdell, E. Fraval, M.~J. Sellars and N.~B.
Manson, Phys. Rev. Lett. 95, 063601 (2005). While in principle the
method should be scalable to light at the quantum level, this has
not been demonstrated yet. Of course, improvements on these
coherence times may be expected in the foreseeable future, however
there is no evidence that these improvements will be significant.

\bibitem{Colbeck}
R. Colbeck, {\em Quantum And Relativistic Protocols For Secure
Multi-Party Computation}, PhD Thesis, University of Cambridge.

\bibitem{RND}
S. Pironio, A. Ac\'{i}n, S. Massar, A. Boyer de la Giroday, D.~N.
Matsukevich, P. Maunz, S. Olmschenk, D. Hayes, L. Luo, T.~A.
Manning and C. Monroe, {\em Random numbers certified by Bell's
theorem} Nature \textbf{464} 1021 (2010).

\bibitem{PM}
S. Pironio and S. Massar,{\em Security of practical private randomness generation},  arXiv:1111.6056.

\bibitem{Fehr}
S. Fehr, R. Gelles and C. Schaffner, {\em Security and
Composability of Randomness Expansion from Bell Inequalities},
arXiv:1111.6052.

\bibitem{Vazirani}
U.~V. Vazirani and T. Vidick, {\em Certifiable Quantum Dice-Or,
testable exponential randomness expansion}, arXiv:1111.6054.


\bibitem{Magniez}
F. Magniez, D. Mayers, M. Mosca and H. Ollivier, {\em Self-Testing
of Quantum Circuits}, Proceedings of 33rd International Colloquium
on Automata, Languages and Programming, volume 4051, series
Lecture Notes in Computer Science, 72 (2006).

\bibitem{singlet}
C.-E. Bardyn, T. C. H. Liew, S. Massar, M. McKague and V. Scarani,
{\em Device independent state estimation based on Bell's
inequalities}, Phys. Rev. A \textbf{80}, 062327 (2009).

\bibitem{MYS12}
M. McKague, T. H. Yang and V. Scarani, {\em Robust self-testing of
the singlet}, arXiv:1203.2976.

\bibitem{bancal}
J.-D. Bancal, N. Gisin, Y. C. Liang and S. Pironio, {\em
Device-Independent Witnesses of Genuine Multipartite
Entanglement}, Phys. Rev. Lett. {\bf 106} 250404 (2011).


\bibitem{BC}
S. L. Braunstein and C. M. Caves, {\em Wringing out better Bell
inequalities}, Ann. Phys. {\bf 202} 22 (1990).

\bibitem{CHSH}
J. F. Clauser, M. A. Horne, A. Shimony and R. A. Holt, {\em
Proposed experiment to test local hidden-variable theories}, Phys.
Rev. Lett. {\bf 23} 880 (1969).

\bibitem{Q set}
M. Navascu\'es, S. Pironio and A. Ac\'{i}n, {\em Bounding the set
of quantum correlations} \emph{Phys. Rev. Lett.} {\bf 98} 010401
(2007); {\em A convergent hierarchy of semidefinite programs
characterizing the set of quantum correlations}, New J. Phys. {\bf
11}, 045021 (2009).

\bibitem{CGLMP}
D. Collins, N. Gisin, N. Linden, S. Massar and S. Popescu, {\em
Bell inequalities for arbitrarily high-dimensional systems}, Phys.
Rev. Lett. {\bf 88} 040404 (2002).




\bibitem{Rphd}
R. Renner, {\em Security of Quantum Key Distribution}, PhD Thesis,
arXiv:quant-ph/0512258.


\bibitem{GPA}
C. H. Bennett, G. Brassard, C. Crepeau and U. M. Maurer, {\em
Generalized privacy amplification}, IEEE Trans. Inf. Th.
\textbf{41}, 6, 1915 (1995).

\bibitem{Azuma}
K. Azuma, {\em Weighted sums of certain dependent random
variables}, Tohoku Mathematical Journal \textbf{19}, 357 (1967).

\bibitem{nprivampl}
R. Arnon-Friedman, E. H\"anggi and A. Ta-Shma, {\em Towards the
Impossibility of Non-Signalling Privacy Amplification from
Time-Like Ordering Constraints},  arXiv:1205.3736; R.
Arnon-Friedman and A. Ta-Shma, {\em On the limits of privacy
amplification against non-signalling memory attacks},
arXiv:1211.1125.




















\end{thebibliography}
\end{document}